\documentstyle[a4,epsf,amstex]{article}
 
\newcommand{\bee}{\begin{equation}}
\newcommand{\ee}{\end{equation}}
\newcommand{\beea}{\begin{eqnarray}}
\newcommand{\eea}{\end{eqnarray}}

\begin{document}
\thispagestyle{empty}
\parskip=12pt
\raggedbottom
 
\def\mytoday#1{{ } \ifcase\month \or
 January\or February\or March\or April\or May\or June\or
 July\or August\or September\or October\or November\or December\fi
%\space\number\day , \\
%\space \number\time , \\
 \space \number\year}
\noindent
\hspace*{9cm} BUTP--98/1\\
%\hspace*{9cm} (DRAFT)\\
\vspace*{1cm}
\begin{center}
{\LARGE The index theorem in QCD with a finite cut-off }
\footnote{Work supported in part by Schweizerischer Nationalfonds,
by Iberdrola, Ciencia y Tecnologia, Espa\~na and by the Ministerio 
de Educacion y Cultura, Espa\~na.}
 
\vspace{1cm}
%\vspace{.5cm}
%\vspace{.5cm}
Peter Hasenfratz, Victor Laliena, 
\\
Institute for Theoretical Physics \\
University of Bern \\
Sidlerstrasse 5, CH-3012 Bern, Switzerland

\vspace{0.5cm}
Ferenc Niedermayer\footnote{On leave from the Institute of Theoretical
Physics, E\"otv\"os University, Budapest}
\\
Paul Scherrer Institute, CH-5232 Villigen PSI, Switzerland

\vspace{0.5cm}
\mytoday \time \\ \vspace*{0.5cm}

\nopagebreak[4]
 
\begin{abstract}
The fixed point Dirac operator on the lattice has exact chiral zero modes on
topologically non-trivial gauge field configurations independently whether
these configurations are smooth, or coarse. 
The relation $n_L-n_R = Q^{\rm FP}$, where $n_L$ $(n_R)$ is the number 
of left (right)-handed zero modes and $Q^{\rm FP}$ is the fixed point 
topological charge holds not only in the continuum limit, but also at 
finite cut-off values. The fixed point action, which is determined by 
classical equations, is local, has no doublers and complies with 
the no-go theorems by being chirally non-symmetric. The index theorem 
is reproduced exactly, nevertheless. In addition, the fixed point Dirac 
operator has no small real eigenvalues except those at zero, i.e. there are no 
'exceptional configurations'.
\end{abstract}
 
\end{center}
\eject
In the continuum the Dirac operator of massless fermions in a smooth 
background gauge field with non-zero topological charge has zero
eigenvalues. The corresponding eigenfunctions are chiral and the number of
left- and right-handed zero modes are related to the topological charge of the
gauge configuration: $n_L-n_R = Q$ \cite{AS}. It is believed by many that this
theorem has important consequences on the low energy properties of QCD. 
A possible intuitive picture of a typical gauge configuration in QCD is that 
of a gas or liquid of instantons and anti-instantons with quantum fluctuations
\cite{DS}. For such configurations, the Dirac operator is expected to have
a large number of quasi-zero modes which could be responsible for spontaneous
chiral symmetry-breaking \cite{BC}.

Typical gauge field configurations contributing to the path integral are not
smooth, however. In addition, the regularization which is necessary to define
the quantum field theory breaks some other conditions of the index theorem
as well. Standard lattice formulations, for example, violate the conditions of
the index theorem in all possible ways: the topological charge of coarse gauge
configurations is not properly defined and the Dirac operator breaks the
chiral symmetry in an essential way. In the continuum limit these problems
disappear, while close to the continuum some trace of the index theorem can be
identified \cite{GH}. In general, however, the expected chiral
zero modes are washed away and, even worse, unwanted real modes occur as
lattice artifacts ('exceptional configurations'
\footnote{We call a configuration exceptional if,
due to some real eigenvalue close to the origin, 
the quark propagator
becomes singular when the bare mass is still distinctly different from its
critical value.} ) \cite{EX}.

We are going to show here that a lattice formulation of QCD working with the
fixed point (FP) action (classically perfect action) %\cite{HN,BB,UW,BO,PH,FN}
[6--11]
and FP topological charge \cite{BBHN,3536} which are defined in the context 
of Wilson's renormalization group theory \cite{KW}, solves all these problems:
the index theorem remains valid on the lattice even if the cut-off is small 
(the resolution is poor). We find it amazing that a theorem which is based 
on differential geometry and topology finds its validity in a context where 
none of these notions seem to be defined. 

The FP action in QCD is determined by classical equations. Many
aspects of these equations and of their solutions have been studied earlier.
In particular, it can be shown that the symmetries of the continuum
theory (infinitesimal translation, rotation, chiral transformation, etc.)
are realized on the {\em solutions} of the FP equations of motion
although the FP lattice action itself is not manifestly invariant \cite{FN}.
Actually, the index theorem on the lattice can be demonstrated by
using these FP equations to connect the lattice problem with that of
the continuum, where the Atiyah-Singer theorem is valid \cite{PH}. Here we
shall follow another way which allows to prove the index theorem on the
lattice directly and leads to additional results as well. 

We shall first give the essential steps of the argument and then present some
details in the second part of the paper.

The fermion part of the FP action can be written as \footnote {We use
dimensionless quantities everywhere. The dimensions are carried by the lattice
unit $a$.}
\begin{equation}
\label{1}
S_{\rm f}^{\rm FP}(\bar{\psi},\psi,U)=
\sum_{n,n'}\bar{\psi}_n h_{n n'}(U)\psi_{n'}\,,
\end{equation}
where $h_{n n'}(U)$ is a specific lattice form of the Dirac operator. 
In eq.~(\ref{1}) the colour and Dirac indices are suppressed. We consider one
flavour, $N_f=1$. 
The FP Dirac operator is local, has no doublers, satisfies 
the hermiticity property
\begin{equation}
\label{2}
h^\dagger=\gamma_5 h \gamma_5 \,,
\end{equation}
and, in complying with the Nielsen-Ninomiya theorem \cite{NN}, it is not
chiral invariant. This is realized, however, in a very special way
--- the chiral symmetry breaking part of the fermion propagator
is given by
\begin{equation}
\label{3}
\frac {1}{2}\{ h^{-1}_{n n'} (U), \gamma_5 \} \gamma_5 = 
R_{n n'}(U) \,,
\end{equation}
where $\{,\}$ denotes the anticommutator.
The hermitian matrix $R_{n n'}(U)$ is trivial in Dirac
space and it is {\it local}. This chiral symmetry breaking term
appears only due to the non-chiral-invariant blocking transformation
of the fermion fields, and not by an ad hoc term added to the action.
Due to the locality of $R$ the chiral properties of the propagating quark are
not affected by this breaking term.
The precise form of $R$ depends on the block
transformation whose FP we are considering. There are block
transformations which give simply $R_{n n'} ={\rm const}\cdot \delta_{n n'}$.
When the averaging extends over the hypercube only (as it is the case for all
the block transformations considered so far), then $n-n'$ extends only over
the hypercube as well,
but in general $R$ has an extension of $O(a)$. 
Eq.~(\ref{3}) with a local $R$ is a highly non-trivial property of 
the FP action. Indeed, $h^{-1}(U)$ is the propagator 
in the background gauge field $U$, hence it is non-local. For a general
case, its chiral symmetry breaking part is expected to be non-local as well.
For the standard Wilson action and for the currently used other 
improved actions $R$ is non-local, indeed.

Ginsparg and Wilson observed a long time ago that eq.~(\ref{3}) with a local
$R$ is the mildest way a local action can break chiral symmetry \cite{GW}. 
They have also demonstrated that the triangle anomaly for $U=1$ 
(free fermions) is correctly reproduced in this case. 
This work remained, however, largely unnoticed since no solution was known 
to the remnant chiral symmetry condition in eq.~(\ref{3}) for QCD. 
The FP action is a solution as we shall demonstrate in 
the second part of this paper (see also in \cite{PH}).

To avoid some singularities at intermediate steps of the calculation
we introduce a small quark mass which will be sent to zero at the end:
\begin{equation}
\label{4}
h_{n n'}(U) \to \hat{h}_{n n'}(U)=h_{n n'}(U) + m_{\rm q}\, \delta_{n n'} \,.
\end{equation}
Due to eq.~(\ref{3}) the chiral symmetry breaking part of $\hat{h}$ satisfies
the relation
\begin{equation}
\label{5}
\frac{1}{2} \{ \hat{h}_{n n'}, \gamma_5 \} \gamma_5 = 
 (h^\dagger R h)_{n n'} +  m_q \, \delta_{n n'} \,.
\end{equation}
This equation will play a basic role in deriving the following results:

{\bf I.} The expectation value of the divergence of the singlet axial
vector current in a background gauge field has the form 
\begin{equation}
\label{8}
\langle \bar{\nabla}_\mu J_\mu^5(n) \rangle =
- \langle \bar{\psi}_n \gamma_5 \left(h^\dagger R h \psi \right)_n + 
\left( \bar{\psi} h^\dagger R h\right)_n \gamma_5 \psi_n \rangle
-2 m_{\rm q} \langle \bar{\psi}_n \gamma_5 \psi_n \rangle \,.
\end{equation}
In eq.~(\ref{8}) the expectation value is defined as
\begin{equation}
\label{9}
\langle {\cal O}(\bar{\psi},\psi,U) \rangle 
= \frac{1}{Z(U)} \int D\bar{\psi} D\psi {\cal O}(\bar{\psi},\psi,U)
\exp\{ -\sum_{m,n} \bar{\psi}_{m} \hat{h}_{m n}(U)\psi_{n} \} \,,
\end{equation}
where $Z(U)$ is given by the analogous integral without the
${\cal O}(\bar{\psi},\psi,U)$ factor. In the $m_{\rm q} \rightarrow 0$ limit
\footnote {The $m_{\rm q} \rightarrow 0$
limit will always be taken even if it is not indicated
explicitly in the following equations.}, the last term in eq.~(\ref{8}) is
dominated by the zero modes (if there are any)
\begin{equation}
\label{91}
%-2 m_{\rm q} \langle \bar{\psi}_n \gamma_5 \psi_n \rangle =
%2\left[ \sum_{i=1}^{n_{\rm R}} \psi_{\rm R}^{(i)*}(n) \psi_{\rm R}^{(i)}(n) 
%- \sum_{j=1}^{n_{\rm L}} \psi_{\rm L}^{(j)*}(n) \psi_{\rm L}^{(j)}(n) 
%\right] \,,
\lim_{m_{\rm q}\to 0} m_{\rm q}
\langle \bar{\psi}_n \gamma_5 \psi_n \rangle =
\sum_{j=1}^{n_{\rm L}} \psi_{\rm L}^{(j)*}(n) \psi_{\rm L}^{(j)}(n)
-\sum_{i=1}^{n_{\rm R}} \psi_{\rm R}^{(i)*}(n) \psi_{\rm R}^{(i)}(n) \,,
\end{equation} 
where $\psi^{(i)}_R(n)$, $\psi^{(j)}_L(n)$ are the normalized 
$\lambda=0$ right- and left-handed eigenfunctions of the eigenvalue problem 
of the Dirac operator
\begin{equation}
\label{51}
\sum_{n'} h_{n n'}(U) \Psi_{n'} = \lambda \Psi_n \,.
\end{equation}

{\bf II.} Summing over $n$ in eq.~(\ref{8}) one obtains
\begin{equation}
\label{151}
0= 2\, {\rm Tr}(\gamma_5 h R) + 2(n_R - n_L)  \, ,
\end{equation} 
where the trace ${\rm Tr}$ is over colour, Dirac and configuration space. 
The first term is a pseudoscalar gauge invariant functional of 
the background gauge field. Further, it can be shown that it is 
a FP operator. On the other hand, its density behaves 
for smooth fields $U$ as
\begin{equation}
\label{14}
- \large\langle \bar{\psi}_n \gamma_5 \left(h^\dagger R h \psi \right)_n + 
\left( \bar{\psi} h^\dagger R h\right)_n \gamma_5 \psi_n \large\rangle
\to
\frac{1}{32\pi^2} \epsilon^{\mu\nu\alpha\beta}
F_{\mu\nu}^{a}(n)F_{\alpha\beta}^{a}(n) 
\end{equation}
up to terms which are of higher order in the vector potential and/or 
derivatives. It follows then
\begin{equation}
\label{150}
{\rm Tr}(\gamma_5 h R) =Q^{\rm FP} \, ,
\end{equation}
where $Q^{\rm FP}$ is the FP topological charge defined earlier in
the Yang-Mills theory \cite{3536}. 
As argued in \cite{BBHN,3536}, $Q^{\rm FP}$ (if it is used together with 
the FP action) defines a correct topological charge: it associates 
a unique integer number $q$ to any configuration \footnote{There exist
configurations whose topological charge might change  under a small 
deformation. This is the case when a small (size $O(a)$) topological object
falls through the lattice. In the saddle point equation, which defines the
fixed point action, there occur two degenerate absolute minima in this
situation. \cite{HN,BBHN}. In the very moment when the topological charge 
changes the
number of zero modes will also change as required by the index theorem.}, 
where $q$ satisfies 
the inequality: action $\ge 8\pi^2 |q|$, with equality for 
the instanton solutions. Eqs.~(\ref{151},\ref{150}) give the index theorem.

\begin{figure}[htb]
\begin{center}
%\vskip 10mm
\leavevmode
\epsfxsize=110mm
\epsfbox{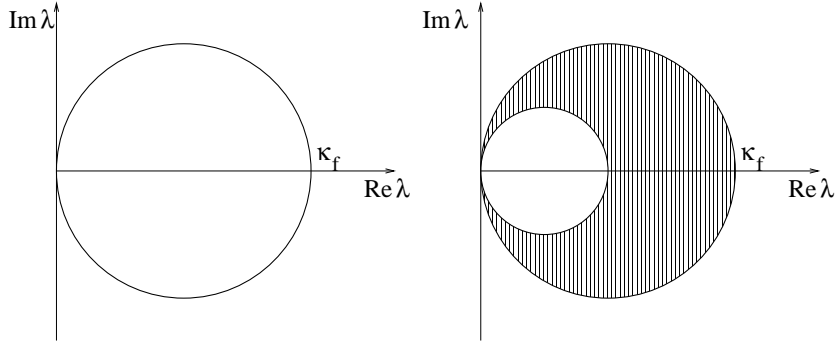}
%\vskip 10mm
\end{center}
\caption{The spectrum of the fixed point Dirac operator:
a) for the special case $R_{nn'}=1/\kappa_{\rm f}\cdot\delta_{nn'}$
the eigenvalues lie on a circle,
b) for a more general set of block transformations the eigenvalues
lie between the two circles.}
\label{fig}
\end{figure}

{\bf III.} Eq.~(\ref{5}) constraints the spectrum $\{ \lambda \}$ of the Dirac
operator. For renormalization group transformations 
('blocking out of continuum' \cite{BOC}) which lead to $R_{n n'}=
1/{\kappa_{\rm f}}\cdot \delta_{n n'}$, where $\kappa_{\rm f}$ is a constant
parameter entering the block transformation,
the eigenvalues $\lambda$ lie on 
a circle whose center is on the real axis in the point $\kappa_{\rm f}/2$ 
and has a radius $\kappa_{\rm f}/2$ (fig.~\ref{fig}).
In general, for a local $R$, the eigenvalues lie between two circles tangent
to the
imaginary axis -- the the circle described previously and a smaller
one, as shown on the second picture in fig.~\ref{fig}. 
This spectrum excludes the existence of exceptional configurations.

We go through the steps now leading to the results above. As usual, the axial
vector current is constructed from the chiral symmetric part of the action,
\begin{equation}
\label{121}
h_{\rm SYM}=\frac{1}{2}[h,\gamma_5]\gamma_5=
\hat{h}-\frac{1}{2}\{\hat{h},\gamma_5 \} \gamma_5 \,.
\end{equation}
Performing a local chiral transformation one obtains for the divergence 
of the singlet axial vector current
\begin{equation}
\label{7}
\bar{\nabla}_\mu J_\mu^5 (n)= 
\bar{\psi}_n \gamma_5 \left( h_{\rm SYM} \psi \right)_n + 
\left( \bar{\psi} h_{\rm SYM} \right)_n \gamma_5 \psi_n \,, 
\end{equation}
where ${\bar \nabla}_\mu f(n)=f(n)-f(n-\hat\mu)$. 
Eqs.~(\ref{7},\ref{121},\ref{5}) give
\begin{eqnarray}
\label{10}
\langle \bar{\nabla}_\mu J_\mu^5 (n) \rangle & = &
\langle \bar{\psi}_n \gamma_5 ( \hat{h}\psi )_n + 
( \bar{\psi} \hat{h})_n \gamma_5 \psi_n \rangle \nonumber \\
 & & - \langle \bar{\psi}_n \gamma_5 \left( h^\dagger R h \psi \right)_n + 
\left( \bar{\psi} h^\dagger R h\right)_n \gamma_5\psi_n \rangle \\
 & & -2 m_{\rm q} \langle \bar{\psi}_n \gamma_5 \psi_n \rangle \,.\nonumber
\end{eqnarray}
The first term on the r.h.s. is proportional to the equation of motion 
$\hat{h}\psi = 0$, the second comes from the remnant chiral symmetry 
condition, while the last term is due to the explicit symmetry breaking.
The contribution of the first term in eq.~(\ref{10}) is zero.
%The following type of identity is used here
Indeed,
\begin{equation}
\label{11}
\langle \bar{\psi}_n \gamma_5 ( \hat{h}\psi )_n \rangle =
-\frac{1}{Z(U)} \int D\bar{\psi} D\psi \, \bar{\psi}_n \gamma_5 
\frac{\delta}{\delta\bar{\psi}_n}
\exp\{ -\sum_{m',n'}\bar{\psi}_{m'} \hat{h}_{m' n'} \psi_{n'}\}\,,
\end{equation}
and partial integration by $\bar \psi$ gives then ${\rm tr}\gamma_5=0$. 
Integrating out the
fermions in the third term in eq.~(\ref{10}) gives 
$2m_q {\rm tr}(\gamma_5{\hat h}^{-1}_{n n})$, where the trace is over colour
and Dirac space.
We insert here 
the closure relation of eigenvectors of the hermitian matrix $H=h\gamma_5$. 
It is easy to show using eq.~(\ref{5}) that the zero eigenvalue 
eigenfunctions of $H$ (if there are any) are chiral and they are 
zero eigenvalue eigenfunctions  of $h$ as well. 
In the $m_q \rightarrow 0$ limit only these eigenfunctions
contribute leading to eq.~(\ref{91}). 

Summing over $n$ in eq.~(\ref{8}), integrating out the fermions in the first
term on the r.h.s. and using eq.~(\ref{91}), one obtains eq.~(\ref{151})
immediately. 

Consider now the derivation of eq.~(\ref{14}). We will show first that
\begin{equation}
\label{141}
-2 m_{\rm q} \langle \bar{\psi}_n \gamma_5 \psi_n \rangle \
\to
\frac{1}{32\pi^2} \epsilon^{\mu\nu\alpha\beta}
F_{\mu\nu}^{a}(n)F_{\alpha\beta}^{a}(n) 
\end{equation}
up to terms which are of higher order in the vector potential and/or
derivatives. Eq.~(\ref{141}) can equivalently be written as
\begin{equation}
\label{16}
\sum_{m,n} m_\alpha n_\beta \left. \left\{
\frac{\delta}{\delta A_\nu^b(0)} \frac{\delta}{\delta A_\mu^a(m)}
\langle 2 m_{\rm q}\bar{\psi}_n \gamma_5 \psi_n 
\rangle  \right\}
\right|_{A=0} 
=\frac{1}{4\pi^2}\epsilon^{\mu\nu\alpha\beta}\delta_{ab}.
\end{equation}
We sketch the derivation only and refer to
\cite{PHF} for the details. Using the notation 
\begin{equation}
\label{17}
i J_\mu^a(m') = 
 - \sum_{m,n} \bar{\psi}_n
\left. \frac{\delta}{\delta A_\mu^a(m')} 
\hat{h}_{m n}(U) \psi_n \right|_{A=0} \,,
\end{equation}
and the definition eq.~(\ref{9}), the l.h.s. of eq.~(\ref{16}) can be 
written as
\begin{equation}
\label{18}
 \sum_{m,n} m_\alpha n_\beta
\left.
\langle \left( -2m_{\rm q}\bar{\psi}_n \gamma_5 \psi_n
\right) J_\mu^a(m) J_\nu^b(0) 
\rangle\right|_{A=0} \,.
\end{equation}
If the derivatives in eq.~(\ref{16}) act on $Z(U)$, or twice on $\hat{h}$, the
contribution can be shown to be zero.
The current in eq.~(\ref{17}) is a correct definition for the colour vector
current: it can be shown that ${\bar \nabla}_\mu J_\mu$ complies with the 
Noether theorem. The current can also be written as
\begin{equation}
\label{19}
J_\mu^a(m) = \sum_{l,l'} \bar{\psi}_{m+l} T^a K_\mu(l,l') \psi_{m+l'} \,,
\end{equation}
where $T^a$ are the colour generators and $K_\mu$ is local.

The matrix element in eq.~(\ref{18}) has been considered in \cite{GW} 
when discussing the anomaly (see
the considerations in \cite{GW} between eqs.~(29-41)). 
It can be shown \cite{PHF} that
$K_\mu$ in eq.~(\ref{19}) satisfies the basic sum rules used in the
derivation of the anomaly in \cite{GW}.
The presence of colour in
eq.~(\ref{18}) gives an extra factor $1/2\cdot\delta_{ab}$ in eq(41) 
in \cite{GW} leading to eq.~(\ref{141}). In \cite{GW}, the authors first 
rewrite $-2m_{\rm q}\bar{\psi}_n \gamma_5
\psi_n$ in eq.~(\ref{18}) using the operator equation for  
$\bar{\nabla}_\mu J_\mu^5 (n)$ and show that only the term coming from the
remnant chiral symmetry condition (proportional to $h^\dagger R h$)
contributes. This gives the result claimed in eq.~(\ref{14}). 

We want to demonstrate now that 
\begin{equation}
\label{21}
Q(U)={\rm Tr}(\gamma_5 h R)
\end{equation}  
is a FP operator as stated  in ${\bf II.}$. 
Further, we have to show that the FP Dirac operator satisfies the
remnant chiral symmetry condition eq.~(\ref{3}). For both problems we need 
the classical equations which determine the FP action. 
The fixed point lies in the $\beta \sim 1/g^2 \rightarrow \infty$ hyperplane. 
In this limit the path integral is reduced to classical saddle point equations.
The gauge field part satisfies
\cite{BB}
\begin{equation}
\label{22}
S_{\rm g}^{\rm FP}(V) = \min_{ \{U\} } \left[ S_{\rm g}^{\rm FP}(U)
+ T_{\rm g}(V,U) \right] \,,
\end{equation}
%where $T_g(V,U)$ defines the local averaging procedure giving the block gauge
%field $V$ in terms of the fine variables $U$. 
where $T_g(V,U)$  defines the block transformation relating the block gauge
field $V$ to a local average of the fine variables $U$.
Denote the minimizing configuration in eq.~(\ref{22}) by $U_{\rm min}(V)$. 
The FP Dirac operator $h$ is defined by the recursion equation
\begin{eqnarray}
\label{27}
& & h(V)_{n_B n_B'} = \kappa_{\rm f}\delta_{n_B n_B'} 
~~~~~~~~~~~~~~~~~~~~~~~~~~~~~~~~~~~~~~~~~~~~~~~~~~~~~~~~~~~~~~~~ \\
& & ~~~ - \kappa_{\rm f}^2 b_{\rm f}^2 \sum_{nn'}
\omega(U_{\rm min})_{n_B n} \left( h(U_{\rm min}) +   
\kappa_{\rm f} b_{\rm f}^2 \omega^{\dagger}(U_{\rm min})
\omega(U_{\rm min}) \right)^{-1}_{n n'}
\omega(U_{\rm min})_{n' n_B'}^{\dagger} \,. \nonumber
\end{eqnarray}
In case $h^{-1}$ is defined, eq.~(\ref{27}) is equivalent to the somewhat
simpler recursion relation \cite{632}
\begin{equation}
\label{23}
h(V)^{-1}_{n_B n_B'} =
\frac{1}{\kappa_{\rm f}} \delta_{n_B n_B'} 
+b_{\rm f}^2 \sum_{n,n'} \omega(U_{\rm min})_{n_B n}
h(U_{\rm min})_{nn'}^{-1}\omega(U_{\rm min})_{n' n_B'}^{\dagger} \,.
\end{equation}
Here $\omega(U)$, which is trivial in Dirac space,
defines the way the block fermion fields are constructed
from the fine fermion fields, while $\kappa_{\rm f}$ is a parameter 
of the block transformation. 
The coarse fields live on a lattice with a lattice unit
$a'=sa$, where $a$ is the lattice unit before the transformation, while $s$ is
the scale of the renormalization group transformation. The factor $b_{\rm f}$ 
is $s^{d_\psi}$, where $d_\psi$ is the engineering dimension of the field
(=3/2), while the averaging function satisfies the normalization condition
$\sum_n \omega(U=1)_{n_B n}=1$.

Eq.~(\ref{23}) (or eq.~(\ref{27})) can be solved by iteration. 
If the scale change $s$ of the
transformation is (infinitely) large, the fixed point can be reached in a
single renormalization group step. In this case, the fine field $U_{min}$ 
lives on
an infinitely fine lattice, and it is smooth. The Dirac operator
$h$ over this field is arbitrarily close to the massless continuum Dirac
operator. Since $\omega$ is trivial in Dirac space, the second term on the
r.h.s. of eq.~(\ref{23}) is chiral invariant, which leads to the remnant
chiral symmetry condition in eq.~(\ref{3}) with 
$R_{n n'}=1/\kappa_{\rm f}\cdot\delta_{n n'}$. 
If $s$ is finite, one can start again on an infinitely fine lattice, 
but the transformation should be iterated. One obtains in every iteration 
step some contribution to the chiral symmetry breaking part of $h^{-1}$, 
and a non-trivial, but local $R$ builds up in this case as claimed 
in eq.~(\ref{3}) before. 
If $h^{-1}$ is not defined (due to zero modes), 
one can use eq.~(\ref{27}) to derive the corresponding remnant symmetry 
condition for $h$ itself with somewhat more algebra.

By adding to the FP action $S^{\rm FP}(U)$ a small perturbation 
$\epsilon {\cal O}(U)$, the path integral 
in the saddle point approximation leads to
the extra contribution 
$\epsilon {\cal O}'(V)=\epsilon {\cal O}(U_{\rm min}(V)$. 
A FP operator is reproduced by the renormalization group 
transformation up to a trivial rescaling related to the dimension 
of the operator:
\begin{equation}
\label{24}
s^{-d_{\cal O}} {\cal O}^{\rm FP}(V) =  
{\cal O}^{\rm FP}\left(U_{\rm min}(V)\right) \,,
\end{equation} 
where $s$ is the scale of the RGT and $d_{\cal O}$ is the dimension of the 
operator ${\cal O}$ \cite{PH}. 
Since the topological charge is a dimensionless quantity, it satisfies
the FP equation
\begin{equation}
\label{25}
Q^{\rm FP}(V)=Q^{\rm FP}\left(U_{\rm min}(V)\right) \,.
\end{equation} 
We want to show that the operator ${\rm Tr}(\gamma_5 h R)$ satisfies 
eq.~(\ref{25}). Using eqs.~(\ref{3},\ref{27}) and the fact that $\omega$ is
trivial in Dirac space, one obtains the following recursion
relation for $R$
\begin{equation}
\label{26}
R(V)_{n_B n_B'} = \frac{1}{\kappa_{\rm f}} \delta_{n_B n_B'} + 
\frac{b_{\rm f}^2}{ \kappa_{\rm f}}
\sum_{n,n'}\omega(U_{\rm min})_{n_B n} R(U_{\rm min})_{n n'}
\omega(U_{\rm min})_{n' n_B'}^{\dagger} \,.
\end{equation} 
With the help of eqs.~(\ref{27},{28}) one obtains after some algebra that
$Q(U)$ in eq.~(\ref{21}) satisfies eq.~(\ref{25}). It is, therefore the FP
topological charge studied earlier in spin and gauge models \cite{BBHN,3536}.

Finally, let us discuss the spectrum of the FP Dirac operator.
Let $\psi$ be a normalized solution of the eigenvalue equation,
$h\psi=\lambda\psi$. Eq.~(\ref{5}) (written in the form
$\frac{1}{2}(h+h^\dagger) = h^\dagger Rh$) implies 
${\rm Re}\lambda=|\lambda|^2(\psi,R\psi)$. 
For the special case $R_{n n'}=1/\kappa_{\rm f}\cdot\delta_{n n'}$
the eigenvalues $\lambda$ lie on a circle of radius 
$\kappa_{\rm f}/2$ touching the imaginary axis, as shown in fig.~\ref{fig}.
In general, the eigenvalues lie between two circles tangent to the imaginary
axis - the circle described previously and a smaller one as shown in
fig.~\ref{fig}. The smaller circle has a finite radius which follows from the
locality of $R$. In particular, one can show that for the block transformation
used in \cite{UW,PK}
$1/\kappa_{\rm f} \le (\psi,R\psi) \le 2/\kappa_{\rm f}$
hence the radius of the smaller circle is $\kappa_{\rm f}/4$.
This property excludes exceptional configurations, appearing e.g. for
the Wilson action and causing serious problems in numerical simulations.
As a consequence of eq.~(\ref{2}) the solutions with 
${\rm Im}\lambda\ne 0$ come in complex conjugate pairs and satisfy
$(\psi,\gamma_5\psi)=0$. Due to eqs.~(\ref{3},\ref{5}), those with 
$\lambda=0$ are chiral eigenstates, $\gamma_5\psi=\pm\psi$.

% &&&&&&&&&&&&&&&&&&&&&&&&&&&&&&&&&&&&&     

\section{Acknowledgements}
The authors are indebted for discussions with C.~B.~Lang, M.~L\"uscher
and H.~Neuberger. P.~H. thanks R.~Crewther for the help to resolve his 
confusion concerning continuum anomaly equations. 
P.~H. is indebted for the warm hospitality at the Departament
d'Estructura i Constituents de la Materia, Barcelona. V.~L. acknowledges the
financial support from Ministerio de Educacion y Cultura, Espa\~na, under
grant PF9573193582.

%\eject
 
\newcommand{\PL}[3]{{Phys. Lett.} {\bf #1} {(19#2)} #3}
\newcommand{\PR}[3]{{Phys. Rev.} {\bf #1} {(19#2)}  #3}
\newcommand{\NP}[3]{{Nucl. Phys.} {\bf #1} {(19#2)} #3}
\newcommand{\PRL}[3]{{Phys. Rev. Lett.} {\bf #1} {(19#2)} #3}
\newcommand{\PREPC}[3]{{Phys. Rep.} {\bf #1} {(19#2)}  #3}
\newcommand{\ZPHYS}[3]{{Z. Phys.} {\bf #1} {(19#2)} #3}
\newcommand{\ANN}[3]{{Ann. Phys. (N.Y.)} {\bf #1} {(19#2)} #3}
\newcommand{\HELV}[3]{{Helv. Phys. Acta} {\bf #1} {(19#2)} #3}
\newcommand{\NC}[3]{{Nuovo Cim.} {\bf #1} {(19#2)} #3}
\newcommand{\CMP}[3]{{Comm. Math. Phys.} {\bf #1} {(19#2)} #3}
\newcommand{\REVMP}[3]{{Rev. Mod. Phys.} {\bf #1} {(19#2)} #3}
\newcommand{\ADD}[3]{{\hspace{.1truecm}}{\bf #1} {(19#2)} #3}
\newcommand{\PA}[3] {{Physica} {\bf #1} {(19#2)} #3}
\newcommand{\JE}[3] {{JETP} {\bf #1} {(19#2)} #3}
\newcommand{\FS}[3] {{Nucl. Phys.} {\bf #1}{[FS#2]} {(19#2)} #3}

\eject


\begin{thebibliography}{99}

\bibitem{AS}
M.~Atiyah and I.~M.~Singer, Ann. Math. 93 (1971) 139;\\
A.~S.~Schwarz,Phys.Lett. 67B (1977) 172;
L.~Brown, R.~Carlitz and C.~Lee, Phys.Rev. D16 (1977) 417.

\bibitem{DS}
E.~V.~Shuryak, Rev.Mod.Phys. 65 (1993) 1;
T.~Sch\"afer and E.~V.~Shuryak, hep-ph/9610451 and
references therein;\\
D.~Diakonov, Prog.Part.Nucl.Phys. 36 (1996) 1.

\bibitem{BC}
T.~Banks and A.~Casher, Nucl.Phys. B169 (1980) 103.

\bibitem{GH}
R.~Narayanan and H.~Neuberger, Phys. Rev. Lett. 71 (1993) 3251 and
Nucl. Phys. B443 (1995) 305;
R.~Narayanan and P.~Vranas, Nucl. Phys. B506 (1997) 373;
R.~Edwards, U.~Heller, R.~Narayanan and R.~Singleton, hep-lat/9711029;
for recent works with an extended list of further references see 
C.~R.~Gattringer, I.~Hip and C.~B.~Lang, Phys. Lett. B409 (1997) 371;
C.~R.~Gattringer and I.~Hip, hep-latt/9712015.

\bibitem{EX}
W.~Bardeen, A.~Duncan, E.~Eichten, G.~Hockney, H.~Thacker,
hep-lat/971084, hep-lat/9705008
%exceptional configurations
 
\bibitem{HN}
P.~Hasenfratz and F.~Niedermayer, Nucl. Phys. B414 (1994) 785;

\bibitem{BB}
T.~DeGrand, A.~Hasenfratz, P.~Hasenfratz and F.~Niedermayer, Nucl. Phys. B454
(1995) 587.

\bibitem{UW}
U.-J.~Wiese, Phy. Lett. B315 (1993) 417;
W.~Bietenholz and U.-J.~Wiese, Nucl. Phys. B(Proc.Suppl.) 34 (1994) 516.

\bibitem{BO}
W.~Bietenholz, R.~Brower, S.~Chandrasekharan and U.-J.~Wiese,
Nucl. Phys. B(Proc.Suppl.) 53 (1997) 921.

\bibitem{PH}
for a recent summary on the perfect actions, see
P.~Hasenfratz, hep-lat/9709110;
for a pedagogical introduction, see 
P.~Hasenfratz, in the proceedings of the 'NATO ASI - Confinement, Duality and
Non-perturbative Aspects of QCD', 1997, Ed. P. van Baal, Plenum Press, 
to be published.

\bibitem{FN}
F.~Niedermayer, Nucl. Phys. B (Proc. Suppl.) 60A (1998) 257.

\bibitem{BBHN}
M.~Blatter, R.~Burkhalter, P.~Hasenfratz and f.~Niedermayer, Phy. Rev. D53
(1996) 923;
R.~Burkhalter, Phys. Rev. D54 (1996) 4121;
W.~Bietenholz, R.~Brower, S.~Chandrasekharan and U.-J.~Wiese, hep-lat/970401.

\bibitem{3536} 
T.~DeGrand, A.~Hasenfratz and T.~Kovacs, hep-lat/9705009;
F.~Farchioni and A.~Papa, hep-lat/9709012.

\bibitem{KW}
K.~Wilson and J.~Kogut, Phys. Rep. C12 (1974) 75;
K.~Wilson, Rev. Mod. Phys. 47 (1975) 773, {\it ibid.},  55 (1983) 583;
K.~Wilson, Adv. Math. 16 (1975) 176, 444.

\bibitem{NN}
N.~B.~Nielsen and M.~Ninomiya, Phys. Lett. B105 (1981) 219; Nucl. Phys. B185
(1981) 20.

\bibitem{GW} 
P.~H.~Ginsparg and K.~G.~Wilson, Phys. Rev. D25 (1982) 2649.

\bibitem{BOC}
W.~Bietenholz and U.-J.~Wiese, Phys. Lett. B378 (1996) 222.

\bibitem{PHF}
P.~Hasenfratz, in the proceedings of 'Advanced School on Non-Perturbative
Quantum Field Physics' (Pe\~niscola) and
of the YKIS'97 'Non-Perturbative QCD -
Structure of the QCD Vacuum', (Kyoto), in preparation.

\bibitem{632}
T.~DeGrand, A.~Hasenfratz, P.~Hasenfratz P.~Kunszt and F.~Niedermayer,
Nucl. Phys. B(Proc.Suppl.) 53 (1997) 942;
W.~Bietenholz and U.-J.~Wiese, Nucl. Phys. B464 (1996) 319.

\bibitem{PK}
P.~Kunszt, hep-lat/9706019, \\
see also the second reference in \cite{PH}.


\end{thebibliography}
\end{document}